\documentclass[twocolumn,twoside]{IEEEtran} 
\usepackage{epsf}

\def\BibTeX{{\rm B\kern-.05em{\sc i\kern-.025em b}\kern-.08em
    T\kern-.1667em\lower.7ex\hbox{E}\kern-.125emX}}

\newcommand{\eg}{e.g.}
\newcommand{\ie}{i.e.}

\begin{document}

\title{Redundant Arrays of IDE Drives} 

\author{D. A. Sanders, {\em Member, IEEE}, L. M. Cremaldi, {\em Member, IEEE}, 
V. Eschenburg, C. N. Lawrence, C. Riley, {\em Member, IEEE}, D. J. Summers, 
D. L. Petravick
\thanks{
This work has been submitted to the IEEE for possible publication. Copyright 
may be transferred without notice, after which this version will be superseded. 
Manuscript submitted to IEEE Transactions On Nuclear Science,
November 25, 2001; revised March 19, 2002. This work was supported in part by 
the U.S. Department of Energy under Grant Nos. DE-FG05-91ER40622 and 
DE-AC02-76CH03000.}
\thanks{D. A. Sanders, L. M. Cremaldi, V. Eschenburg, C. N. Lawrence, 
C. Riley, and D. J. Summers are with the
University of Mississippi, Department of Physics and Astronomy, 
University, MS 38677 USA (telephone: 662-915-5438, e-mail: 
sanders@phy.olemiss.edu.)}
\thanks{D. L. Petravick is with the
Fermi National Accelerator Laboratory, CD-Integrated Systems 
Development, MS 120, Batavia, IL 60510-0500 USA (telephone: 
630-840-3935, e-mail: petravick@fnal.gov)}
\thanks{Publisher Item Identifier 10.1109/TNS.2002.801699}
}

\maketitle

\begin{abstract}
The next generation of high-energy physics experiments is expected to 
gather prodigious amounts of data. New methods must be developed to handle 
this data and make analysis at universities possible. We examine 
some techniques that use recent developments in commodity hardware. We 
test redundant arrays of integrated drive electronics (IDE) disk drives for 
use in offline high-energy physics data analysis. IDE redundant array 
of inexpensive disks (RAID) prices now equal the cost 
per terabyte of million-dollar tape robots!  The arrays can be scaled to 
sizes affordable to institutions without robots and used when fast random 
access at low cost is important. We also explore three methods of moving data 
between sites; internet transfers, hot pluggable IDE disks in FireWire cases, 
and writable digital video disks (DVD-R).
\end{abstract}


\section{Introduction}

\PARstart{W}{e} report tests, using the Linux operating system, of redundant 
arrays of integrated drive electronics (IDE) disk drives for use in particle 
physics Monte Carlo simulations and data analysis \cite{farm,farm2}. Parts 
costs of total systems using commodity IDE disks are now at the \$4000 per 
terabyte level. A revolution is in the making. Disk storage prices have now 
decreased to the point where they equal the cost per terabyte of 300 terabyte 
Storage Technology tape silos. The disks, however, offer far better 
granularity; even small institutions can afford to deploy systems. The faster 
random access of disk versus tape is another major advantage. Our tests 
include reports on software redundant arrays of inexpensive disks -- Level 5 
(RAID-5)  systems running under Linux 2.4 using Promise Ultra 
100 disk controllers. RAID-5 protects data in case of a 
catastrophic single disk failure by providing parity bits. Journaling file 
systems are used to allow rapid recovery from system crashes. We also report 
on using FireWire (IEEE 1394) to PCI (Peripheral Component Interconnect) 
interfaces. FireWire PCI cards allow 
sixty-three devices (\eg ~a combination of computers and disks) per 
card. The maximum Firewire bus speed is currently limited to 400 megabits
per second. FireWire is also hot pluggable.

Our data analysis strategy is to encapsulate data and CPU processing power 
together. Data is stored on many PCs. Analysis of a particular part of a 
data set takes place locally on, or close to, the PC where the data resides. 
The network backbone is only used to put results together. 
If the I/O overhead is moderate and analysis tasks need more than one 
local CPU to plow through data, then each of these disk arrays could be 
used as a local file server to a few computers sharing a local ethernet switch. 
These commodity 8-port gigabit ethernet switches would be combined with a 
single high end, fast backplane switch allowing the connection of a thousand 
PCs. We have also successfully tested using Network File System (NFS) software 
to connect our disk arrays to computers that cannot run Linux 2.4. 

We examine three ways of moving data between sites; internet transfers, 
hot pluggable IDE disks in FireWire cases, and writable digital video disks 
(DVD-R). Writable 4.7 GB DVD-R disks are now available for \$5. They can be                                                        
read by \$60 DVD-ROM drives and written by the \$500 Pioneer DVR--A03 
drive \cite{pioneer}. 

RAID \cite{RAID} stands for Redundant Array of Inexpensive Disks.  Many 
industry offerings meet all of the qualifications except the inexpensive part, 
severely limiting the size of an array for a given budget. This may change. 
The different RAID levels can be defined as follow:
\begin{itemize}
\item RAID-0: ``Striped.'' Disks are combined into one physical device where 
reads and writes of data are done in parallel. Access speed is fast but there 
is no redundancy.
\item RAID-1: ``Mirrored.'' Fully redundant, but the size is limited to 
the smallest disk.
\item RAID-4: ``Parity.'' For $N$ disks, 1 disk is used as a parity bit and 
the remaining $N-1$ disks are combined. Protects against a single disk 
failure but access speed is slow since you have to update the parity disk for 
each write.
\item RAID-5: ``Striped-Parity.'' As with RAID-4, the effective size 
is that of $N-1$ disks. However, since the parity information is also 
distributed evenly among the $N$ drives the bottleneck of having to update 
the parity disk for each write is avoided. Protects against a single disk 
failure and the access speed is fast.
\end{itemize}

RAID-5, using enhanced integrated drive electronics (EIDE) disks under Linux 
software, is now available \cite{RAID5}. Redundant disk arrays do provide 
protection in the most likely single disk failure case, that in which a 
single disk simply stops working. This removes a major obstacle to building 
large arrays of EIDE disks. However, RAID-5 does not totally protect against 
other types of disk failures. RAID-5 will offer limited protection in the 
case where a single disk stops working but causes the whole EIDE bus to fail 
(or the whole EIDE controller card to fail), but only temporarily stops them 
from functioning. This would temporarily disable the whole RAID-5 array. 
If replacing the bad disk solves the problem, i.e.~the failure did not 
permanently damage data on other disks, then the RAID-5 array would recover 
normally. Similarly if only the controller card was damaged then replacing 
it would allow the RAID-5 array to recover normally. However, if more 
than one disk was damaged, especially if the file or directory 
structure information was damaged, the entire RAID-5 array would be 
damaged. The remaining failure mode would be for a disk to be delivering 
corrupted data. There is no protection for this inherent to RAID-5; however, 
a longitudinal parity check on the data, such as a checksum record count 
(CRC), could be built into event headers to flag the problem. 
Redundant copies of data that are very hard to recreate are still needed. 
RAID-5 does allow one to ignore backing up data that is only moderately 
hard to recreate.

\section{Large Disks}
\begin{figure}[ht!]
\centerline{\epsfxsize 3.5 truein \epsfbox{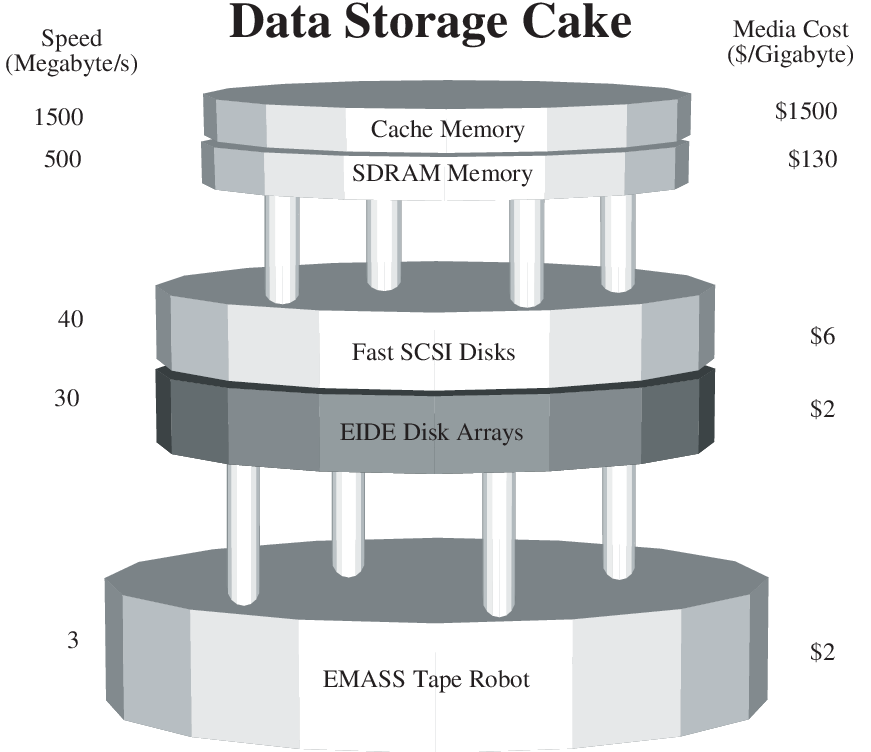}}
\caption{\hangindent=0pt
Historically the speed and cost of data storage has increased as one moved 
from tape to disk to RAM. EIDE RAID-5 disk arrays add another layer to the 
data storage cake.  One doesn't have to worry as much about tape backup 
except for data that is very hard to recreate. The chance of losing data 
is lower than with plain scratch disks.  The cost of EIDE RAID-5 is close 
to that of tape robots and the random access speed of disk is much faster.}
\label{Cake}
\end{figure}

In today's marketplace, the cost per terabyte of disks with EIDE interfaces 
is about a third that of disks with SCSI (Small Computer System Interface), 
as illustrated in Fig. \ref{Cake}. 
The EIDE interface is limited to 2 drives on each bus and SCSI is 
limited to 7 (14 with wide SCSI). The only major drawback of EIDE disks is 
the limit in the length of cable connecting the drives to the drive 
controller. This limit is nominally 18 inches; however, we have successfully 
used 24 inch long cables \cite{star}. 
Therefore, one is limited to 10 disks per box for an array (or perhaps 20 
with a ``double tower''). To get a large RAID array one needs to use large 
capacity disk drives. There have been some problems with using large disks, 
primarily the maximum addressable size. We have addressed these problems in 
an earlier paper \cite{CHEP98}. Because of these concerns and because we 
wanted to put more drives into an array than could be supported by the 
motherboard we opted to use PCI disk controller cards. We tested both 
Promise Technologies ULTRA 66 \cite{Promise66} and ULTRA 
100 \cite{Promise100} disk controller cards, which each support 
four drives.

Using arrays of disk drives, as shown in Table \ref{Disks}, the cost per 
terabyte is similar to that of cost of Storage Technology tape silos. 
However, RAID-5 arrays offer a lot better granularity since they are scalable 
down to a terabyte. For example, if you wanted to store 10 TB of data you 
would still have to pay about \$1,000,000 for the tape silo but only 
\$40,000 for a RAID-5 array. Thus, even small institutions can afford to 
deploy systems. Therefore, as seen in Fig. \ref{Cake}, ``you can have your 
cake and eat it too''. 

\section{RAID Arrays}
There exist disk controllers that implement RAID-5 protocols right in the 
controller, for example 3ware's Escalade 7850  \cite{3ware,3ware2}, which 
will handle up to eight EIDE drives. These controllers cost \$600 and did 
not support disk drives larger than 137 Gigabytes \cite{Kent}; so we focused 
our attention on software RAID-5 implementations \cite{RAID5,HOWTO}, which 
we tested extensively.

\subsection{Hardware}
We have examined both Maxtor 
DiamondMax \cite{maxtor,maxtor540,maxtor100} and IBM 
DeskStar \cite{IBM75} hard disks. 
For RAID-5 the disk 
partitions must be all of the same size. The only trouble we 
had was when Maxtor changed the capacity for the 80 GB disk from 81.9 GB 
to 80 GB. We had to repartition the 81.9 GB disks to 80 GB (plus a wasted 
partition of 1.9 GB). Fortunately this happened to a test array and 
not while trying to replace a failed disk in a working RAID-5 array. 
Disk manufacturers have recently decided to define one GB as 1000 MB, 
rather than 1024 MB. The drives we consider for use with a RAID-5 array are 
compared in Table \ref{Disks}. In general, the internal I/O speed of a disk 
is proportional to its rotational speed and increases as a function of 
platter capacity. 
\begin{table}[htb]
\caption{Comparison of Large EIDE Disks for a RAID-5 Array}
\begin{center}
{\tt
\tabcolsep=0.85mm
\begin{tabular}{lrcccr}\hline
&&&&& Spin-Up\\
&&&Cost& GB per&Current\\
Disk Model& GB&RPM&per GB&platter&at 12V\\  \hline
Maxtor D540X\cite{maxtor540}& 80& 5400& \$2.11& 20& 2.00A\\
Maxtor D536X\cite{maxtor100}& 100& 5400& \$2.27& 33& 0.64A\\
&&&&&\\
Maxtor D540X\cite{maxtor540}& 160& 5400& \$1.85& 40& 1.80A\\
IBM 75GXP \cite{IBM75}& 75& 7200& \$3.19& 15& 2.00A\\
IBM 120GXP\,\cite{IBM120} & 120 & 7200& \$2.91& 40& 2.00A\\
\hline
\end{tabular}
}
\end{center}
\label{Disks}
\end{table}

When assembling an array we had 
to worry about the ``spin-up'' current draw on the 12V 
part of the power supply. With 8 disks in the array (plus the system 
disk) we would have exceeded the capacity of the power supply that came 
with our tower case, so we decided to add a second off-the-shelf power 
supply rather than buying a more expensive single supply. 
By using 2 power supplies we benefit from 
under loading the supplies. The benefits include both a longer 
lifetime and better cooling since the heat generated is distributed 
over 2 supplies, each with their own cooling fans. 
We used the hardware shown in Table \ref{Configuration} for our array test. 
Many of the components we chose are generic; many components from 
other manufacturers also work. 
We have measured the wall power consumption for the whole disk array box in 
Table \ref{Configuration}. It uses 276 watts at startup and 156 watts during 
normal sustained running.

\begin{table}[ht!]
\begin{center}
\caption[]{700\,GB RAID-5 Configuration}
\label{Configuration}
\tabcolsep=1.5mm
\renewcommand{\arraystretch}{1.10}
\vspace*{3pt}
\begin{tabular}{lr} \hline
System      & Unit \\
 Component  & Price \\ \hline
100GB Maxtor system disk \cite{maxtor100}            & \$227 \\
8 -- 100GB Maxtor RAID-5 disks \cite{maxtor100}      & \$227 \\
2 -- Promise ATA/100 PCI cards \cite{Promise100}     &  \$27 \\
4 -- StarTech 24" ATA/100 cables \cite{star}         &   \$3 \\
AMD Athlon 1.4 GHz/266 CPU \cite{AMD}                & \$120  \\
Asus A7A266 motherboard, audio \cite{asus}           & \$132  \\
2 -- 256MB DDR PC2100 DIMMs                          & \$35   \\
In-Win Q500P Full Tower Case \cite{inwin}            & \$77   \\
Sparkle 15A @ 12V power supply  \cite{sparkle}       & \$34 \\
2 -- Antec 80mm ball bearing case fans               &  \$8  \\
110 Alert temperature alarm \cite{cool}              & \$15 \\
Pine 8MB AGP video card \cite{pine,nVIDIA}           & \$20 \\
SMC EZ card 10/100 ethernet \cite{smc,RealTek}       & \$12 \\
Toshiba 16x DVD, 48x CDROM                           & \$54 \\
Sony 1.44 MB floppy drive                            & \$12 \\
KeyTronic 104 key PS/2 keyboard                      & \$7  \\
DEXXA 3 button PS/2 mouse                            & \$4   \\
\cline{2-2}
\hfill Total                                         & \$2682  \\
\hline
\end{tabular}
\end{center}
\end{table}

To install the second power supply we had to modify our tower 
case with a jigsaw and a hand drill. We also had to use a jumper to 
ground the green wire in the 20-pin block ATXPWR connector to fake 
the power-on switch.

When installing the two disk controller cards care 
had to be taken that they did not share interrupts with other highly 
utilized hardware such as the video card and the ethernet card. We 
also tried to make sure that they did not share interrupts with each 
other. There are 16 possible interrupt requests (IRQs) that allow the 
various devices, such as EIDE controllers, video cards, mice, serial, 
and parallel ports, to communicate with the CPU. Most PC operating 
systems allow sharing of IRQs but one would naturally want to avoid 
overburdening any one IRQ. There are also a special class of IRQs 
used by the PCI bus, they are called PCI IRQs (PIRQ). Each PCI card 
slot has 4 interrupt numbers. This means that they share some IRQs 
with the other slots; therefore, we had to juggle the cards we used 
(video, 2 EIDE controllers, and an ethernet). 

When we tried to use a disk as a ``Slave'' on a motherboard 
EIDE bus, we found that it would not run at the full speed of the bus 
and slowed down the access speed of the entire RAID-5 array. This was 
a problem of either the motherboard's basic input/output system (BIOS) 
or EIDE controller. This problem was not in evidence when using the disk 
controller cards. Therefore, we decided that rather than take a factor 
of 10 hit in the access speed we would rather use 8 instead of 9 hard disks. 

\subsection{Software}
For the actual tests we used Linux kernel 2.4.5 with the RedHat 7 
(see http://www.redhat.com/) distribution (we had to upgrade the kernel 
to this level). The latest stable kernel version is 2.4.18 
(see http://www.lwn.net/). We needed the 2.4.x kernel to allow full support 
for ``Journaling'' file systems. Journaling file systems provide rapid 
recovery from crashes. A computer can finish 
its boot-up at a normal speed, rather than waiting to perform a file 
system check (FSCK) on the entire RAID array. This is then conducted 
in the background allowing the user to continue to use the RAID array. 
There are now 4 different Journaling file systems: XFS, a port from 
SGI \cite{XFS}; JFS, a port from IBM \cite{JFS}; ext3 \cite{ext3}, 
a Journalized version of the standard ext2 file system; and ReiserFS 
from namesys \cite{ReiserFS}. Comparisons of these Journaling file 
systems have been done elsewhere \cite{Journaling}. 
When we tested our RAID-5 arrays only ext3 and the ReiserFS were 
easily available for the 2.4.x kernel; therefore, we tested 2 different 
Journaling file systems; ReiserFS and ext3. We opted on using ext3 for 
two reasons 1) At the time there were stability problems with ReiserFS 
and NFS (this has since been resolved with kernel 
2.4.7) and 2) it was an extension of the standard ext2fs (it was 
originally developed for the 2.2 kernel) and, if synced properly could 
be mounted as ext2. Ext3 is the only one that will allow direct upgrading 
from ext2, this is why it is now the default for RedHat 7.2.

NFS is a very flexible system that allows one to manage files on several 
computers inside a network as if they were on the local hard disk. So, 
there's no need to know what actual file system they are stored under nor 
where the files are physically located in order to access them. Therefore, 
we use NFS to connect these disks arrays to computers that cannot run 
Linux 2.4. We have successfully used NFS to mount this disk array on the 
following types of computers: a DECstation 5000/150 running Ultrix 4.3A, 
a Sun UltraSparc 10 running Solaris 7, a Macintosh G3 running MacOS\,X, and 
various Linux boxes with both the 2.2 and 2.4 kernels. We are currently 
using two of these RAID-5 boxes to run analysis software with the 
{\sc BaBar} KANGA code and the CMS CMSIM/ORCA code.

We have performed a few simple speed tests. The first was \hbox{``hdparm 
-tT /dev/xxx''}. This test simply reads a 64 MB chunk of data and 
measures the speed. On a single drive we saw read/write speeds of about 
30 MB/s. On the whole array we saw a drop to 28 MB/s. When we 
tried writing a text file using a simple FORTRAN program (we wrote ``All 
work and no play make Jack a dull boy'' $10^{8}$ times), the speed 
was 22.34 MB/s \footnote{Since we originally submitted 
this paper we have tested a new Asus motherboard (the A7M266 
with the AMD 761 North Bridge chip) and 
got significant increases in speed for the RAID-5 array.} While mounted 
via NFS over 100 Mb/s ethernet the speed was 2.12 MB/s, limited by 
both the ethernet speed and communication overhead.
In the past \cite{farm2}, we have been able to get a much higher 
fraction of the rated ethernet bandwidth by using the lower level TCP/IP 
socket protocol \cite{TCPIP} in place of the higher level NFS protocol.  
TCP/IP sockets are more cumbersome to program, but are much faster.

We also tested what actually happens when a disk fails by turning the power 
off to one disk in our RAID-5 array.  One could continue to read and write 
files, but in a ``degraded'' mode, that is without the parity safety net. 
When a blank disk was added to replace the failed disk, again one could 
continue to read and write files in a mode where the disk access speed is 
reduced while the system rebuilt the missing disk as a background job. 
This speed reduction in disk access was due to the fact that the parity 
regeneration is a major disk access in its own right. For more details, 
see reference \cite{HOWTO}.

The performance of Linux IDE software drivers is improving. The latest 
standards \cite{t13} include support for command overlap, READ/WRITE direct 
memory access QUEUED commands, scatter/gather data transfers without 
intervention of the CPU, and elevator seeks. Command overlap is a protocol 
that allows devices that require extended command time to perform a bus 
release so that commands may be executed by the other device on the bus. 
Command queuing allows the host to issue concurrent commands to the same 
device. Elevator seeks minimize disk head movement by optimizing the order 
of I/O commands.

We did encounter a few problems. We had to modify ``MAKEDEV'' to allow for 
more than eight IDE devices, that is to allow for disks 
beyond ``/dev/hdg''. For version 2.x one would have to actually modify the 
script; however, for version 3.x we just had to modify the file 
``/etc/makedev.d/ide''.

Another problem was the 2 GB file size limit. Older operating system and 
compiler libraries used a 32 bit ``long-integer'' for addressing files; 
therefore, they could not normally address files larger than 2 GB ($2^{31}$). 
There are patches to the Linux 2.4 kernel and glibc but there are still some 
problems with NFS and not all applications use these patches. 

We have found that the current underlying file systems (ext2, ext3, reiserfs) 
do not have a 2 GB file size limit.  The limit for ext2/ext3 is in the 
petabytes. The 2.4 kernel series supports large files (64-bit offsets). 
Current versions of GNU libc support large files. However, by default the 
32-bit offset interface is used. To use 64-bit offsets, C/C++ code must be 
recompiled with the following as the first line:
\begin{verbatim}
#define _FILE_OFFSET_BITS 64
\end{verbatim}
or the code must use the *64 functions (i.e. open becomes open64, etc.) 
if they exist. This functionality is not included in GNU FORTRAN (g77); 
however, it should be possible to write a simple wrapper C program to 
replace the OPEN statement (perhaps called open64). We 
have succeeded in writing files larger than 2 GB using a simple C 
program with ``\#define \_ FILE\_ OFFSET\_ BITS 64'' as the 
first line. This works over NFS version 3 but not version 2. 

While RAID-5 is recoverable for a hardware failure, there is no 
protection against accidental deletion of files. To address this 
problem we suggest a simple script to replace the ``rm'' command. 
Rather than deleting files it would move them to a ``/raid/Trash'' or 
better yet a ``/raid/.Trash'' directory on the RAID-5 disk array 
(similar to the ``Trash can'' in the Macintosh OS). The system 
administrator could later purge them as space is needed using an algorithm 
based on criteria such as file size, file age, and user quota.

\section{FireWire}
FireWire was developed by Apple and is an IEEE standard (IEEE 1394) 
defining a high speed serial bus. This bus is also named ``i.Link'' by 
Sony. It is referred to as IEEE 1394 or just 1394 in the Linux world 
\cite{linux1394}. It is a serial bus similar in principle to the 
Universal Serial Bus (USB), but runs at speeds of up to 400 Mb/s and is 
intended to replace the SCSI bus; however, it is not centered around a PC 
(\ie  ~there may be none or multiple PCs on the same bus). The FireWire 
bus allows up to sixty-three devices per chain. Also, because it has a 
mode of transmission which guarantees bandwidth, it is used for digital 
video cameras and similar devices. In general it is hot swappable.

There are 2 main chipsets supported under Linux. The supported chipsets are
Texas Instruments PCILynx/PCILynx2 and OHCI compliant chips (produced by 
various companies). FireWire drivers are now included in RedHat and other 
distributions and are supported in the 2.4.x kernel (with patches for 
the 2.2.x kernel). However, not all drivers are included in a standard 
installation nor is it a default option when upgrading the kernel. The 
driver for storage devices, such as hard disks (SBP-2) , was not included 
in kernels until the 2.4.7 kernel. For these reasons, we are including the 
basic instructions here.

We got FireWire working on a Linux box by following the following steps:
\begin{enumerate}
    
\item We used an inexpensive PCI FireWire controller, for a cost of \$25. 
It was an OHCI-1394 card with a VIA controller.  

\item The kernel used was Linux 2.4.12 as released by Linus Torvalds and 
Alan Cox's -ac3 patch. Alan's patches can be downloaded at 
http://www.bz2.us.kernel.org/pub/linux /kernel/people/alan/linux-2.4/.
The -ac series is basically what Red Hat and other distributions base their 
kernels on, and includes drivers not in stock 2.4.12.

\item We had to enabled FireWire support when configuring the kernel. 
This involved turning on the following:
\begin{verbatim}
IEEE 1394 (FireWire) support (EXPERIMENTAL)
  OHCI-1394 support
  SBP-2 support (Harddisks etc.)
\end{verbatim}
(The RAWIO driver is not necessary for storage devices.  In addition,
you will need the SCSI disk driver enabled in the kernel, even if you
don't have a real SCSI interface on the machine. This is because 
FireWire is treated as a SCSI channel.)

\item After rebooting with the new kernel, some recent distributions 
should detect the FireWire card and install the correct drivers. If not, 
the following modules need to be manually loaded, in this order: 
\begin{verbatim}

ohci1394 
sbp2

\end{verbatim}
The sbp2 driver is somewhat finicky; it helps to have a few seconds 
delay between the two modprobes. The command ``cat /proc/scsi/scsi'' 
should list the attached storage devices (disks, CD-ROMs, etc.):
\begin{verbatim}
Attached devices: 
Host: scsi1 Channel: 00 Id: 00 Lun: 00
  Vendor: Maxtor  Model: 1394 storage   Rev: 60  
  Type:   Direct-Access  ANSI SCSI revision: 02
\end{verbatim}
Some of the output may not make sense if an IDE-FireWire (1394) bridge is 
in use; we noticed the non-Maxtor drive had strange output. 

\end{enumerate}

At the moment, the devices are added in more-or-less random order.
The only way to guarantee ordering is to manually hot-plug them.  We
don't know if this is a software limitation or an artifact of the
plug\&play nature of FireWire (there's no permanent ID setting like IDE or
SCSI have).  Presumably if one writes a volume header label (\eg  ~with
\hbox{tune2fs -L}) to each disk you could get around this problem. 

Hot plugging seems to work with the following caveat. Do not unplug 
a FireWire device without unmounting it first. While you do not have 
to shutdown the computer to remove the device you do have to unmount 
it. Once unmounted, disconnect the device physically and then run 
``\hbox{rescan-scsi-bus.sh -r}''. For new devices, plug them in and run 
``\hbox{rescan-scsi-bus.sh}''. The script can be downloaded at 
http://www.garloff.de/kurt/linux/ \hbox{rescan-scsi-bus.sh} 

We successfully configured two FireWire disks, after formatting 
the disks using ext2, (but any common file system, such as ext3 or 
RieserFS, would work) 
as a RAID-5 array. One of the disks used the new Oxford 911 FireWire 
to EIDE interface chip \cite{oxsemi,oxsemi2,oxsemi3,OWC}. We have succeeded 
in writing a DVD-R using the Pioneer DVR-A03 over FireWire. 

\section{High Energy Physics Strategy}
\subsection{Data Storage Strategy -- Event Persistence}
We encapsulate data and CPU processing power. A block of real or Monte 
Carlo simulated data for an analysis is broken up into groups 
of events and distributed once to a set of RAID disk boxes, which each 
may also serve a few additional processors via a local 8-port gigabit ethernet 
switch \footnote{
{\tabcolsep=0mm
\begin{tabular}[t]{lclcr}
D--Link & \hspace{2mm} & DGS--1008T 8-port gigabit ethernet switch & & \$765 \\
Linksys & & EG0008 8-port gigabit ethernet switch     & & \$727 \\
Netgear & & GS508T 8-port gigabit ethernet switch     & & \$770 \\
Netgear & & GS524T 24-port gigabit ethernet switch    & & \$1860 \\
D-Link  & & DGE500T RJ45 gigabit ethernet PCI adapter & \hspace{2mm} & \$46 \\
\end{tabular}} \hfill \break
(See http://www.dlink.com/\,, http://www.linksys.com/products/\,, \hfill \break  
~and http://www.netgear.com/)} 
Dual processor boxes would also add more local CPU power. Events are kept 
physically contiguous on disks to minimize I/O. Events are only built once. 
Event parallel processing has a long history of success in high energy 
physics \cite{farm,farm2,E769,E769b,Kunz}. The data from each analysis are 
distributed 
among all the RAID arrays so all the computing power can be brought to bear 
on each analysis. For example, in the case of an important analysis (such 
as a Higgs analysis), one could put 50 GB of data onto each of 100 RAID arrays 
and then bring the full computing power of 700 CPUs into play. Instances of 
an analysis job are run on each local cluster in parallel. Several analyses 
jobs may be running in memory or queued to each local cluster to level loads. 
The data volume of the results (\eg ~histograms) is small and is gathered 
together over the network backbone. Results are examined and the analysis is 
rerun. The system is inherently fault tolerant. If three of a hundred clusters 
are down, one still gets 97\% of the data and analysis is not impeded. 

RAID-5 arrays should be treated as fairly secure, large, high-speed ``scratch 
disks''. RAID-5 just means that disk data will be lost less frequently. 
Data which is very hard to re-create still needs to reside on tape. The 
inefficiency of an offline tape vault can be an advantage. Its harder to erase 
your entire raw data set with a single keystroke, if thousands of tapes have 
to be physically mounted. Someone may ask why all the write protect switches 
are being reset before all is lost. Its the same reason the Air Force has real 
people with keys in ICBM silos.

The granularity offered by RAID-5 arrays allows a university or 
small experiment in a laboratory to set up a few terabyte computer farm, 
while allowing a large Analysis Site or Laboratory to set up a few 
hundred terabyte or a petabyte computer system. For a large site, they 
would not necessarily have to purchase the full system at once, but 
buy and install the system in smaller parts. This would have two 
advantages, primarily they would be able to spread the cost over a 
few years and secondly, given the rapid increase in both CPU power and 
disk size, one could get the best ``bang for the buck''.

What would be required to build a 300 terabyte system (the same size  
as a tape silo)? Start with eight 160\,GB Maxtor disks in a box. The 
Promise Ultra133 card allows one to exceed the 137\,GB limit \footnote{
Promise Technology's Ultra133 TX2 PCI 
controller card uses a wider 48-bit data address versus the older 28-bit 
address, which is limited to $2^{28}$~512 byte blocks or 137 Gigabytes. 
The card controls four disks and has a \$59 list price. 
(See http://www.promise.com/\hfill \break 
marketing/datasheet/file/Ultra133tx2DS.pdf)}. 
Each box provides 
7 $\times $ 160\,GB = 1120\,GB of usable RAID-5 disk space in addition to 
a CPU for computations. 300 terabytes is reached with 270 boxes. Use 40 
commodity 8-port gigabit ethernet switches (\$800 each) to connect the 
270 boxes to a 40-port, high end, fast backplane ethernet switch 
\cite{lucent,lucent2}. 
This could easily fit in a room that was formerly occupied by a few old 
Mainframes, say an area of about a hundred square meters. The power consumption 
would be 42 kilowatts. One would need to build up operational experience for 
smooth running.  As newer disks arrive that hold yet more data, even a petabyte 
system would become feasible.

\subsection{Data Transfer Strategy}
For small amounts of data and to update analysis software one can use 
internet file transfers, preferably via ``rsync''. The program ``rsync'' 
remotely copies files and  uses a remote-update protocol to greatly speedup 
file transfers when the destination file already exists. This remote-update 
protocol allows ``rsync'' to transfer just the differences between two sets 
of files across the network link, using an efficient checksum-search 
algorithm. Some of the additional features of ``rsync'' are: 
support for copying links, devices, owners, groups and permissions; 
can use any transparent remote shell, including ``rsh'' or ``ssh''; 
can tunnel over encrypted connections and is compatible with Kerberized 
rsh/ssh authentication; 
and does not require root privileges. The only problem is the 
available bandwidth. Internet2 may ameliorate this problem but given the 
prevalence of Napster-like programs competing with data transfers, 
this is not a certainty. The other method would be to use some form of 
removable, and universally readable media. Two new methods are hot 
pluggable IDE disks in \$90 FireWire cases \cite{OWC}, and DVD-R disks. 
Since FireWire works on Linux, Windows 98SE, and Macintosh OS9 and 
OSX, one can use hot pluggable EIDE disks in FireWire cases as a simple 
method of transferring reasonable amounts of data or even full sets of 
analysis software. In any case, its best not to try and transfer any chunk 
of data more than once.  Local CPUs and disks are far less expensive than 
wide area networks. 

Writable 4.7 GB DVD-R disks can be purchased for \$5.  They can be read 
by \$60 DVD-ROM drives and written by the \$500 Pioneer DVR-A03 
drive \cite{pioneer}. Linux is capable of writing DVD-Rs. However, 
the software to do so is not available under a free license. It is an 
enhanced version of ``cdrecord'', the free program that writes CDs, CD-Rs, 
and CD-RWs. A demo version that will write up to 1 GB is available from 
the author's FTP site \cite{cdrecord}. An alternative, which is free, 
is to use the patch for cdrecord \cite{dvdpatch}. Using this patched 
version of ``cdrecord'', we have succeeded in writing a DVD-R using the 
Pioneer DVR-A03 both internally (it's an EIDE device) and over FireWire. 
The specific kernel used was linux 2.4.18 plus the pre1 patch from 
Marcelo Tosatti \cite{Marcelo,Marcelo2}, the pre1-ac2 patch from Alan 
Cox \cite{Alan}, and the ieee1394 tree \cite{linux1394}. 
We used a patched version of cdrecord 1.11a11. The image was a standard 
iso9660 filesystem image created with ``mkisofs'', including a 2880 kB boot 
image.  (The DVD itself contains a complete copy of the February 27, 2002 
snapshot of Debian Linux's upcoming 3.0 release, which would normally take 
up six 700 MB CD-Rs.)  The image took approximately 25 minutes to write at 2x 
speed. The long-term reliability of DVD-R media still needs to be explored.

\section{Conclusion}
We have tested redundant arrays of IDE disk drives for use in 
offline high energy physics data analysis and Monte Carlo simulations. 
Parts costs of total systems using commodity IDE disks are now at 
the \$4000 per terabyte level, the same cost per terabyte as Storage 
Technology tape silos. The disks, however, offer much better granularity; 
even small institutions can afford them. The faster access of disk versus 
tape is a major added bonus. We have tested software RAID-5 systems 
running under Linux 2.4 using Promise Ultra 100 disk 
controllers. RAID-5 provides parity bits to protect data in case of a single 
catastrophic disk failure. Tape backup is not required for data that can be 
recreated with modest effort. Journaling file systems permit rapid recovery 
from crashes. Our data analysis strategy is to encapsulate data and CPU 
processing power. Data is stored on many PCs. Analysis for a particular part 
of a data set takes place locally on the PC where the data resides. The 
network is only used to put results together. Commodity 8-port gigabit 
ethernet switches combined with a single high end, fast backplane switch 
would allow one to connect a thousand PCs, each with a terabyte of disk space. 
Some tasks may need more than one CPU to go through the data even on 
one RAID array.  For such tasks dual CPUs and/or several boxes on one local 
8-port ethernet switch should be adequate and avoids overwhelming the backbone 
switching fabric connecting an entire installation.  Again the backbone 
is only used to put results together. 
We successfully performed simple tests of three methods of 
moving data between sites; internet transfers, hot pluggable EIDE disks in 
FireWire cases, and DVD-R disks.

Current high energy physics experiments, like {\sc BaBar} at SLAC, feature 
relatively low data acquisition rates, only 3 MB/s, less than a third of 
the rates taken at Fermilab fixed target experiments a decade ago 
\cite{farm,farm2}. The Large Hadron Collider experiments CMS and Atlas, 
with data acquisition rates starting at 100 MB/s, will be more challenging 
and require physical architectures that minimize helter skelter data movement 
if they are to fulfill their promise. In many cases, architectures designed 
to solve particular processing problems are far more cost effective than 
general solutions \cite{farm,farm2,E769,E769b}. Some of the techniques 
explored in this paper, to physically encapsulate data and CPUs together, 
may be useful.

\bibliographystyle{IEEE}

\begin{thebibliography}{99}
\bibitem{farm}
S. Amato, ~J.~R.~T. de Mello Neto, ~J. de Miranda, ~C. James, ~D.~J. Summers, 
~and ~S.~B. Bracker, 
\newblock ``The E791 parallel architecture data acquisition system,'' 
\newblock {\em Nucl. Instr. Meth.}, vol. A324, pp. 535--542, Jan.~1993, 
\newblock arXiv:hep-ex/0001003. 

\bibitem{farm2}
S. Bracker, ~K. Gounder, ~K. Hendrix, ~and ~D. Summers, 
\newblock ``A simple multiprocessor management system for event parallel
  computing,'' 
\newblock {\em IEEE Trans. Nucl. Sci.}, vol. 43, pp. 2457--2464, Oct.~1996, 
\newblock arXiv:hep-ex/9511009.

\bibitem{pioneer}Pioneer. 
\newblock (2002) DVR-A03. [ONLINE]. Available: \hfill \break
http://www2.pioneer-eur.com/products/multimed/optical/  \hfill \break
dvrA03.htm 

\bibitem{RAID}
D.~Patterson, G.~Gibson, and R.~Katz, 
\newblock ``A case for redundant arrays of inexpensive disks,'' in 
{\em Proc. 1988 ACM SIGMOD Conf. Management of Data}, vol.~17, 
SIGMOD Record, Chicago, IL, June 1988, pp.~109--116.

\bibitem{RAID5}
M. de~Icaza, I. Molnar, and G. Oxman, 
\newblock ``The linux raid-1,4,5 code,'' in  
\newblock 3rd Annu. Linux Expo'97, Apr. 1997.

\bibitem{star}StarTech.com. 
\newblock (2002) 24 In. Dual Drive Ultra ATA/66/100 \hfill \break
Cable. [ONLINE]. Available: http://www.startech.com/ \hfill \break
ststore/itemdetail.cfm?product{\_}id=IDE66{\_}24 \hfill \break
The ATA/100 standard uses an 80 wire cable to transmit up to 
100 Megabytes per second.

\bibitem{CHEP98}
D. Sanders, C. Riley, L. Cremaldi, D. Summers, and D. Petravick, 
\newblock ``Working with arrays of inexpensive eide disk drives,'' 
\newblock in {\em Proc. Int. Conf. Computing in High- 
Energy Physics (CHEP 98)}, Chicago, IL, Aug. 31 - Sep 4 1998, 
\newblock arXiv:hep-ex/9912067.

\bibitem{Promise66}Promise Technologies,Inc. 
\newblock (1998) Ultra66\,--\,PCI Card for Ultra\hfill \break
ATA/66 Drives. [ONLINE]. Available: http://www.promise.com\hfill \break
/marketing/datasheet/file/Ultra66.pdf

\bibitem{Promise100}Promise Technologies, inc. 
\newblock (2001) Ultra100 TX2 -- Ultra ATA/100 
Controller for 66MHz PCI Motherboards. [ONLINE].  \hfill \break
Available: http://www.promise.com/marketing/datasheet/file/ 
U100{\_}TX2{\_}DS.pdf \hfil \break
\newblock Each ATA/PCI Promise card controls four disks.

\bibitem{3ware}
S. Sanchez. 
\newblock (2001) 3wareÕs new ESCALADE storage switch 
extends performance lead over traditional SCSI RAID controllers 
by 300\%. [ONLINE]. Available: http://www.3ware.com/news/ \hfill \break 
7450{\_}7850.pdf
 
\bibitem{3ware2}
D. Graas. (2001) Escalade Storage Switch. 
[ONLINE]. Available: 
http://www.3ware.com/products/EscaladeReturnLetter.asp

\bibitem{Kent}
K. Abendroth, ``personal communication,'' \hfill \break
email: kent.abendroth@3ware.com.

\bibitem{HOWTO}
J. \O stergaard. 
\newblock (2000) 
\newblock The software-RAID \hfill \break
HOWTO. [ONLINE].
\newblock Available: 
http://www.linuxdoc.org/ \hfill \break
\hbox{HOWTO/Software-RAID-HOWTO.html}.

\bibitem{maxtor}Maxtor.
\newblock (2000) DiamondMax 80. [ONLINE]. \hfill \break
\newblock Available: http://www.maxtor.com/en/documentation/ \hfill \break
data{\_}sheets/diamondmax{\_}80{\_}ata100{\_}datasheet.pdf

\bibitem{maxtor540} Maxtor.
\newblock (2001) DiamondMax D540X. [ONLINE]. \hfill \break
\newblock Available: http://www.maxtor.com/en/documentation/ \hfill \break
data{\_}sheets/d540x{\_}datasheet.pdf 

\bibitem{maxtor100}Maxtor.
\newblock (2001) DiamondMax 536DX. [ONLINE]. \hfill \break
\newblock Available: http://www.maxtor.com/en/documentation/ \hfill \break
data{\_}sheets/536DX{\_}datasheet.pdf 

\bibitem{IBM75}IBM. 
\newblock (2000) IBM Deskstar 75GXP and Deskstar 40GV
hard disk drives. [ONLINE]. Available: \hfill \break
http://www.storage.ibm.com/hdd/prod/deskstar75gxp40gv.pdf

\bibitem{IBM120}IBM. 
\newblock (2001) IBM Deskstar 120GPX hard disk drives. 
[ONLINE]. Available: \hfill \break 
http://www.storage.ibm.com/hdd/desk/ds120gxp.pdf

\bibitem{AMD}AMD.
\newblock (2002) AMD Athlon Processor Product Brief. 
[ONLINE]. Available: http://www.amd.com/us-en/Processors/ \hfill \break
ProductInformation/0,,30{\_}118{\_}756{\_}759\symbol{94}1151,00.html 
\hfill \break
We bought our AMD CPU boxed with a fan.

\bibitem{asus}ASUS. 
\newblock (2002) ASUS A7A266. [ONLINE]. Available: \hfill \break
http://www.asus.com/mb/socketa/a7a266/overview.htm

\bibitem{inwin}In-Win Development, inc. 
\newblock (2002) IW-Q500 ATX Full Tower Case. [ONLINE]. Available:\hfill \break
\hbox{http://www.in-win.com/framecode/ino{\_}q500.html}  \hfill \break
\newblock Note: the Q500P case comes with a 300 Watt power supply.

\bibitem{sparkle}Sparkle Power Inc. 
\newblock (2002) FSP300-60BTV For P4 and Athlon. [ONLINE]. 
Available:\hfill \break
http://www.sparklepower.com/pdf/FSP300-60BTV.pdf \hfill \break
\newblock The Sparkle FSP300-60BTV is used as a second 300 watt supply.
At 12 volts it gives 15 amps.

\bibitem{cool}PC Power \& ~Cooling, Inc. 
\newblock (2000) 110 ALERT Computer Over-Temperature Alarm. [ONLINE]. 
Available:  \hfill \break
http://www.pcpowercooling.com/pdf/110Alert{\_}ds.pdf

\bibitem{pine}PINE Technology. 
\newblock (2001) VGA Card Specification. [ONLINE]. Available:\hfill \break
 http://www.pinegroup.com/pdf/S2/L1/product{\_}list.pdf

 \bibitem{nVIDIA}
 nVIDIA. (2002) nVIDIA Consumer Desktop Solutions. \hfill \break 
 [ONLINE]. Available: http://www.nvidia.com/\hfill \break 
 docs/lo/962/SUPP/NV{\_}LC{\_}02.05.02B.pdf \hfill \break
 The Pine 8MB AGP NVIDIA VANTA LT video card is used to run a
 monitor for diagnostics.

\bibitem{smc}SMC Networks. 
\newblock (2001) EZ Card 10/100. [ONLINE]. 
 Available: http://www.smc.com/drivers{\_}downloads/library/ \hfill \break
 SMC1211.pdf
 
\bibitem{RealTek}
 D. Becker. (1999) A RealTek RTL8129/8139 Fast 
 Ethernet driver for Linux. [ONLINE]. Available: \hfill \break
 http://www.smc.com/drivers{\_}downloads/library/rtl8139.c \hfill \break
 The SMC 1211TX EZ PCI Card uses an rtl8139 10/100 ethernet 
 software driver.
\bibitem{XFS}SGI. 
\newblock (2001) XFS: A high-performance journaling file system. [ONLINE].
\newblock Available: http://oss.sgi.com/projects/xfs/

\bibitem{JFS}
S. Best.
\newblock (2002) Journaled File System Technology for Linux. 
[ONLINE]. Available: \hfill \break
\newblock http://www-124.ibm.com/developerworks/oss/jfs/

\bibitem{ext3}A. Morton.
\newblock (2002) ext3 for 2.4. [ONLINE]. Available: \hfill \break
http://www.zip.com.au/$\sim $akpm/linux/ext3/

\bibitem{ReiserFS}
H. Reiser.
\newblock (2001) Three reasons why ReiserFS is great for you. [ONLINE].
\newblock Available: http://www.reiserfs.org/

\bibitem{Journaling}
R. Galli. 
\newblock (2001) 
\newblock Journal File Systems in Linux.
\newblock {\em Upgrade}. \hfill \break 
\newblock [ONLINE]. Vol. II(6), pp. 1-8, Available: \hfill \break
http://www.upgrade-cepis.org/issues/2001/6/up2-6Galli.pdf

\bibitem{TCPIP}
S. Feit, 
\newblock {\em TCP/IP: Architecture, Protocols, and Implementation}. 
NewYork: McGraw--Hill, 1993.

\bibitem{t13}
P. McLean.
\newblock (2002) Technical Committee T13 AT Attachment. [ONLINE].
\newblock Available: http://www.t13.org/

\bibitem{linux1394}
A. ~Bombe, ~S. ~Rougeaux, ~D. ~Dennedy, ~and~ E. ~Pirker. \hfill \break
\newblock (2002) IEEE 1394 for Linux. [ONLINE]. Available: \hfill \break
\newblock http://linux1394.sourceforge.net/

\bibitem{oxsemi}Oxford Semiconductor, Ltd. 
\newblock (2000) Second generation 1394 to ATA bridge chip 
brings faster, more feature rich PC and MAC storage peripherals. 
[ONLINE]. Available: \hfill \break
http://www.oxsemi.com/press/dec00/index.html

\bibitem{oxsemi2}
\newblock Oxford Semiconductor, Ltd. 
\newblock (2001) OXFW911 IEEE1394 to ATA/ATAPI Native Bridge 
Datasheet. [ONLINE]. Available: \hfill \break
http://www.oxsemi.com/products/IEEE1394/fw911ds.pdf

\bibitem{oxsemi3}Oxford Semiconductor, Ltd. (2000) 
 Customer perception of performance in IEEE 1394 Applications. [ONLINE]. 
 Availabe: http://www.oxsemi.com/products/an/oxan1.pdf

\bibitem{OWC}
For a 5.25" Firewire Disk Case using the Oxford 911 chip see: \hfill \break
\newblock Other World Computing. (2002) OWC Mercury 5.25" 
FireWire External Case. [ONLINE]. Available: \hfill \break
http://eshop.macsales.com/\hfill \break
Catalog{\_}Item.cfm?ID=1445{\&}Item=casekitfw

\bibitem{E769}
For a description of the first UNIX farm at Fermilab see: \hfil \break
C.~Stoughton and D.~J. Summers, 
\newblock ``Using multiple RISC processors in parallel to study charm 
quarks,'' 
\newblock {\em Comput.~Phys.} vol. 6, pp. 371--376, July/Aug.~1992, 
\newblock arXiv:hep-ex/0007002

\bibitem{E769b}
C.~Gay and S.~Bracker, ``The E769 multiprocessor based data acquisition 
system,'' {\em IEEE Trans. Nucl. Sci.}, vol. 34, pp. 870--872, Aug.~1987.

\bibitem{Kunz}
P. F.~Kunz, R. N.~Fall, M. F.~Gravina, J.~H. Halperin,
L.~J.~Levinson, G.~J.~Oxoby, and Q.~H.~Trang,
\newblock ``Experience using the 168/E microprocessor for offline 
data analysis,'' 
\newblock {\em IEEE Trans.~Nucl.~Sci.} vol.~27, pp.~582--586, Feb.~1980.

\bibitem{lucent}
For example, we use the Lucent Cajun P550 Switch with a 23 Gigabits 
per second backplane capacity. Up to six cards may be installed in this 
switch. One option is a card with 48 full duplex 10/100 Base-TX ports. 
See: \hfill \break
Lucent Technologies. (1999) 
\newblock Cajun ~P550 ~Gigabit Switch. \hfill \break 
[ONLINE]. Available: \hfill \break
http://lucent.netlabs.net/ins/products/p550.html 

\bibitem{lucent2}
J. H. Ricciardone and S. Loudermilk. (1999) 
\newblock Lucent Technologies unveils 48-port 10/100 MBPS module 
for Cajun P550 Switch. [ONLINE]. Available: \hfill \break 
http://www.lucent.com/press/0899/990818.nsc.html

\bibitem{cdrecord}
J. Schilling.
\newblock (2001) CDRecord. [ONLINE].
\newblock Available: \hfill \break
ftp://ftp.fokus.gmd.de/pub/unix/cdrecord/ProDVD/

\bibitem{dvdpatch}
N. Mihalache.
\newblock (2001) DVD support for cdrecord. \hfill \break
[ONLINE]. Available: \hfill \break
http://www.abcpages.com/$\sim $mache/cdrecord-dvd.html

\bibitem{Marcelo}
M. Tosatti.
\newblock (2002) 
\newblock The Linux Kernel Archives, \hfill \break
linux-2.4.18. [ONLINE].
\newblock Available: http://www.kernel.org/ \hfill \break
pub/linux/kernel/v2.4/linux-2.4.18.tar.bz2 

\bibitem{Marcelo2}
M. Tosatti. (2002) 
\newblock The Linux Kernel Archives, \hfill \break
patch-2.4.19-pre1. [ONLINE].
\newblock Available: \hfill \break
http://www.kernel.org/pub/linux/kernel/v2.4/testing/ \hfill \break
patch-2.4.19-pre1.bz2

\bibitem{Alan}
A. Cox.
\newblock (2002) 
\newblock The Linux Kernel Archives,  \hfill \break
patch-2.4.19-pre1-ac2. [ONLINE].
\newblock Available: \hfill \break 
http://www.kernel.org/pub/linux/kernel/people/alan/ \hfill \break
patch-2.4.19-pre1-ac2.bz2

\end{thebibliography}

\end{document}